\RequirePackage{fix-cm}
\documentclass{svjour3}                     
\smartqed  
\usepackage{graphicx}
\usepackage{amssymb} 
\usepackage{amsmath} 
\usepackage{wasysym}
\usepackage{color}         
\usepackage{natbib}
\usepackage[pdftitle={Habitability of the early Earth: Liquid water under a faint young Sun facilitated by strong tidal heating due to a closer Moon}, colorlinks = true, breaklinks = true, citecolor = blue, linkcolor = blue, urlcolor = blue, pdfauthor = {Ren\'{e} Heller}]{hyperref}

 \journalname{Pal{\"a}ontologische Zeitschrift (PalZ)}

\begin{document}

\title{Habitability of the early Earth: Liquid water under a faint young Sun facilitated by strong tidal heating due to a closer Moon 
}

\titlerunning{Habitability of the early Earth}        

\author{Ren{\'e} Heller
             \and
            Jan-Peter Duda
            \and
             Max Winkler
             \and
            Joachim Reitner
            \and
            Laurent Gizon
}


\institute{R.~Heller \at
              Max Planck Institute for Solar System Research, Justus-von-Liebig-Weg 3, 37077 G\"ottingen, Germany \\
              Institut f{\"u}r Astrophysik G{\"o}ttingen, Georg-August-Universit{\"a}t G{\"o}ttingen, Friedrich-Hund-Platz 1, 37077 G{\"o}ttingen, Germany\\
              \email{heller@mps.mpg.de}
              \and
              J.-P.~Duda \at
              Department of Geosciences, University of T{\"u}bingen, 72074 T{\"u}bingen, Germany \\
              G\"ottingen Academy of Sciences and Humanities, 37073 G\"ottingen, Germany\\
              \email{jan-peter.duda@geo.uni-tuebingen.de}
              \and
              M.~Winkler \at
              Institut f{\"u}r Mineralogie, Universit{\"a}t M{\"u}nster, Corrensstra{\ss}e 24, 48149 M{\"u}nster, Germany\\
              \email{max.winkler@wwu.de}
              \and
              J.~Reitner \at
              G\"ottingen Centre of Geosciences, Georg-August-Universit{\"a}t G{\"o}ttingen, 37077 G\"ottingen, Germany \\
              G\"ottingen Academy of Sciences and Humanities, 37073 G\"ottingen, Germany\\
              \email{jreitne@gwdg.de}
              \and
              L.~Gizon \at
              Max Planck Institute for Solar System Research, Justus-von-Liebig-Weg 3, 37077 G\"ottingen, Germany \\
              Institut f{\"u}r Astrophysik G{\"o}ttingen, Georg-August-Universit{\"a}t G{\"o}ttingen, Friedrich-Hund-Platz 1, 37077 G{\"o}ttingen, Germany\\
              Center for Space Science, NYUAD Institute, New York University Abu Dhabi, Abu Dhabi, UAE\\
              \email{gizon@mps.mpg.de}
}

\date{Draft \today}

\maketitle

\begin{abstract}
Geological evidence suggests liquid water near the Earth’s surface as early as 4.4 gigayears ago when the faint young Sun only radiated about 70\,\% of its modern power output. At this point, the Earth should have been a global snowball if it possessed atmospheric properties similar to those of the modern Earth. 
An extreme atmospheric greenhouse effect, an initially more massive Sun, release of heat acquired during the accretion process of protoplanetary material, and radioactivity of the early Earth material have been proposed as reservoirs or traps for heat.
For now, the faint-young-sun paradox persists as an important problem in our understanding of the origin of life on Earth. Here we use the constant-phase-lag tidal theory to explore the possibility that the new-born Moon, which formed about 69 million years (Myr) after the ignition of the Sun, generated extreme tidal friction -- and therefore heat -- in the Hadean and possibly the Archean Earth.
We show that the Earth-Moon system has lost ${\sim}3~{\times}~10^{31}$\,J (99\,\% of its initial mechanical energy budget) as tidal heat. Tidal heating of ${\sim}10\,{\rm W\,m}^{-2}$ through the surface on a time scale of 100\,Myr could have accounted for a temperature increase of up to $5^\circ$C on the early Earth. This heating effect alone does not solve the faint-young-sun paradox but it could have played a key role in combination with other effects.
Future studies of the interplay of tidal heating, the evolution of the solar power output, and the atmospheric (greenhouse) effects on the early Earth could help in solving the faint-young-sun paradox.
\end{abstract}

\section{Introduction}
\label{sec:introduction}

Geological records, such as oxygen isotope ($\delta^{18}$O) data from zircons, show that liquid water was present on the Earth's surface as early as 4.4 billion years ago (gigayears ago, Ga)\footnote{Throughout this manuscript we use two time scales, one of which is typically used in astrophysics and one of which is often used in Earth sciences. The first one is the time after formation of the Sun, measured in units of millions of years (Myr) or billions of years (Gyr). The second one is the time before present, with units abbreviated as ``Ga'' for giga-annum \citep{Arndt2011}. We use Ga for ``gigayears ago'' or ``gigayears old''. For the conversion between the time scales we use a solar age of $4.567\,(\pm\,0.06)$\,Gyr determined from $^{204}$Pb/$^{206}$Pb vs. $^{207}$Pb/$^{206}$Pb isotope measurements in the Allende meteorite \citep{Amelin2007}.} \citep{Mojzsis2001,Wilde2001,2014NatGe...7..219V}. This water would not have been pure water near room temperature and pressure but a high-temperature ($300$\,K--$450$\,K) and high-pressure (${\sim}500$\,bar) mix of H$_2$O and CO$_2$. \citet{Liu2004} suggested that CO$_2$ was then removed continuously over ${\sim}100$\,Myr to form carbonate rocks, although this is unfortunately not documented in the geological rock record. Life could have been present as early as 3.8\,Ga to 3.5\,Ga as suggested by possible biosignatures (particularly organic matter and its carbon isotopic composition) preserved in ancient rocks \citep{Schidlowski1988,Rosing1999}, and diverse aquatic life was certainly established by 3.5\,Ga - 3.4\,Ga as for instance evidenced by organic biosignatures and fossil microbial mats \citep{Lowe1980,Walter1980,Allwood2006,vanKranendonk2011,Duda2016,Duda2018,HickmanLewis2018,Homann2019}. $^{18}$O/$^{16}$O isotope ratios in marine cherts and carbonates suggests Archean ocean temperatures between $50^\circ$C and $85^\circ$C \citep{Knauth2005}, although longevity and extent of these conditions remain unclear \citep[see e.g. discussion in][]{Sengupta2020}. Moreover, no evidence for glaciations exist until about 2.9\,Ga \citep{vonBrunn1993,Young1998,VanKranendonk2012}.

\begin{figure}
\centering
\includegraphics[width=1.0\linewidth]{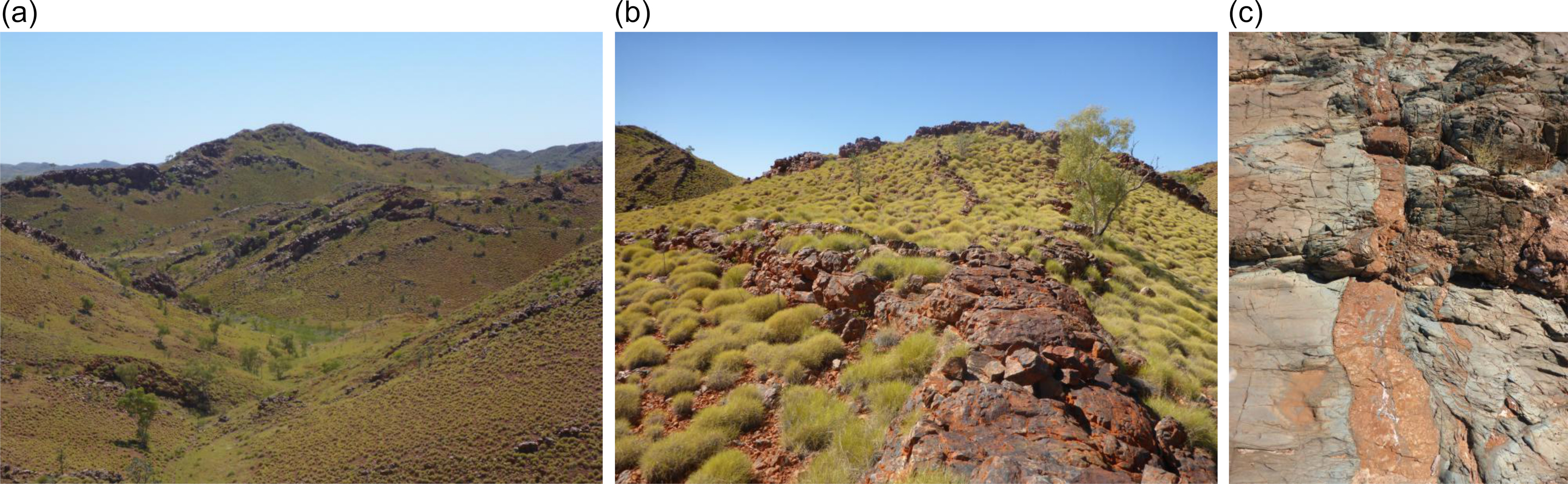}
\caption{(a) Early Archean basaltic rocks in the Pilbara region (Western Australia). These rocks evidently formed in aquatic environments as evidenced by pillow structures. Furthermore, pillow basaltic rocks such as the 3.49\,Ga North Star Basalt are typically crosscut by numerous chert veins, which record intensive hydrothermal pumping of surface water through the crust at that time \citep[``hydrothermal pump hypothesis'';][]{Duda2018}. (b) Detailed outcrop pattern of black chert veins. The vein in the front has a diameter of about two to three meters. (c) Remarkably, these strata also show other types of evidence for brittle deformation such as widespread carbonate cemented fractures such as this 3.47\,Ga Mount Ada Basalt. The brown color of the carbonate is characteristic for Fe-rich dolomites (ankerite). It appears plausible that such features are the product of strong tidal forces exerted by the much closer moon.}
\label{fig:chert_veins}
\end{figure}

Early Archean basalts, such as the 3.49\,Ga North Star Basalt in the Pilbara region (Western Australia) \citep{Hickman1977}, typically exhibit pillow structures (Fig.~\ref{fig:chert_veins}). Since pillow structures form when hot, basaltic magma penetrates into a body of liquid water, they provide direct geological evidence for the presence of liquid surface water on the early Earth\footnote{The formation of pillow basaltic rocks is a widespread phenomenon in today's oceans and been observed and documented at numerous locations \citep[most famously on Hawaii, see e.g.][pp. 94-95, pp. 274-275]{Grotzinger2007}. Importantly, this process is confined to subaquatic surface areas since it requires liquid water and space for the pillows to develop. For these reasons, we consider pillow basalts as robust geological indicators for the presence of liquid water in early-Earth surface environments as well.}. The North Star Basalt and other Paleoarchean pillowed basalts in this area (e.g., Mount Ada Basalt, Apex Basalt, Euro Basalt) are typically crosscut by numerous chert veins, which record hydrothermal pumping of surface water through early Earths crust \citep[``hydrothermal pump hypothesis'';][]{Duda2018}.

As conclusive as the geological record is in regard to the presence of liquid water at (or near) the surface during the first billion years after the formation of the Earth, there is an astrophysical prediction that is in strong disagreement with these observations. Until about 3.5\,Ga, the solar illumination on Earth was only about 70\,\% of its current value. With the solar radiation being the major energy source by far, simple energy balance calculations show that the reduction of the solar luminosity by 30\,\% compared to its modern value should have led to a global snowball Earth for at least the first billion years or so \citep{Sagan1972,1988SciAm.258b..90K}. 

One possibility to prevent global glaciation is through hugely enhanced atmospheric CO$_2$ levels. Levels of 100 to 1000 times the present value have been shown theoretically to prevent a global snowball under a faint Sun 2.5\,Gyr ago \citep{1983Natur.301...53K,Kasting1987}. This would correspond to CO$_2$ partial pressures of between 0.03\,bar and 0.3\,bar, compared to $5~\times~10^{-4}$\,bar today. Even more massive amounts of CO$_2$ with up to between 10\,bar and 100\,bar of a CO$_2$-CO dominated atmosphere have been suggested on the Hadean Earth (4.4\,Ga - 4\,Ga) using model calculations \citep{2001Natur.409.1083N}. The lower end of this range overlaps with theoretical estimates of between 1\, bar \citep{1959Sci...130..245M} and 10\,bar \citep{1983Natur.302..518W} from atmosphere evolution models and with estimates for the CO$_2$ inventory that could have been stored in the lithosphere \citep{1997Sci...276.1217S}.

Alternatively, a 1\,bar atmosphere with a $10^{-5}$ volume mixing ratio of NH$_3$ could have generated the required magnitude of a strong greenhouse effect \citep{Sagan1972}. NH$_3$, however, would be photodissociated through solar UV radiation on a time scale of just about 40\,yr \citep{1979Icar...37..207K} if it were not continuously supplied in adequate quantities by abiotic processes \citep{1982JGR....87.3091K}. Photodissociative destruction of NH$_3$ could also have been prevented by an UV-opaque high-altitude haze composed of organic solids that were produced from CH$_4$ photolysis \citep{1997Sci...276.1217S}. Moreover, CH$_4$ itself could have acted as an efficient greenhouse gas on the early Earth \citep{2008AsBio...8.1127H}. This argument, however, shifts the problem to the production of CH$_4$. Where did it come from? If the only viable source of large amounts of CH$_4$ is through methanogenic microorganisms \citep{2000JGR...10511981P,2002Sci...296.1066K}, then this is a chicken-and-egg dilemma: life would be required to create and maintain the conditions for it to have emerged in the first place.

Other explanations to solve the faint-young-sun paradox include the assumption of a much more massive Sun 4.5\,Ga \citep{1995JGR...100.5457W}, a much lower fractional land coverage on Earth during the Archean \citep{Flament2008}, and different cloud distributions than today, all of which would affect the global albedo \citep{Goldblatt2011}.

The atmospheric composition of the early Earth and the possibility of a greenhouse effect that is much stronger than the contemporary warming of $+33^\circ$C compared to an airless body (or of $+20.3^\circ$C compared to a gray atmosphere model, see Sect.~\ref{sec:balance}) could certainly have played important roles in preventing a global snowball Earth. A specific interplay of various chemical, geothermal, and possibly biotic effects (at least once life was present) could solve the problem.

There is, however, a more apparent effect that has, to our perception, been underappreciated \citep{2012RvGeo..50.2006F} or even overlooked \citep{2020arXiv200606265C,CatlingZahnle2020} in the discussion of the faint-young-sun paradox: that is, geothermal heating induced by the tidal deformation of the Earth by the newly formed Moon. In comparison to the other mechanisms detailed above, tidal heating always has a warming  effect, whereas changes in the Earth's atmospheric composition, cloud coverage, land-to-surface fraction, albedo etc. could in principle also yield even lower temperatures in the absence of empirical model constraints. The energy budget difference between the initial and modern states of the Earth-Moon system provide firm constraints on the total amount of heat that must have been dissipated. Here we present the basic calculations to address this tidal energy budget.

\section{Earth surface temperature}

\subsection{Sources of energy on the early Earth}

The surface temperature of the early Earth was determined by a balance between the energy flux provided by different energy sources and the energy loss through radiation as infrared emission into space. The most important energy sources were absorption of the solar radiation, the translation of the kinetic energy of accreted material (planetesimals) into heat, radiogenic decay of short-lived isotopes in the Earth's mantle and crust, and, to a smaller amount, conversion of gravitational energy into heat during chemical stratification.

The solar radiation varied substantially during the first ${\sim}50$\,Myr after the formation of the Sun (see Sect.~\ref{sec:solar}). After the Moon-forming impact some $69\,(\pm10)$\,Myr after the beginning of the solar system \citep{Maltese2019}, however, the solar flux settled at about 70\,\% of its modern value and then increased at a rate of about 6\,\% per billion years. The accretion of the Earth converted the gravitational energy of the incoming debris into heat on a time scale of 10\,Myr to 100\,Myr, the total amount of which was comparable to the solar heat absorbed at the time \citep{Turcotte1985}. The most important short-lived radionuclides were $^{26}$Al and $^{60}$Fe \citep{Urey1955,Chaussidon2007}. Their short half life times of 0.7\,Myr and 1.5\,Myr, respectively, made them important sources of energy for the first 10\,Myr or so, but negligible afterwards. The radiogenic decay of long-lived isotopes $^{40}$K, $^{232}$Th, $^{235}$U, and $^{238}$U provided about $0.2\,{\rm W\,m}^{-2}$ to the Hadean Earth \citep{1991Icar...90..222S,2005AsBio...5..100G}. For comparison, the modern mean surface output of $^{238}$U and $^{232}$Th radioactive decay is $0.022^{+0.015}_{-0.010}\,{\rm W\,m}^{-2}$ \citep{Shimizu2015} and the global mean heat flow is $0.092\,({\pm\,0.004})\,\mathrm{W/m}^2$ \citep{Davies2010}. The residual heat is mostly due to cooling of the Earth from its accretion.

The Moon might have played another important role in the early Earth's energy budget. The key physical effect here is tidal heating, a phenomenon that is well known from observations of some of the rocky and icy moons of the solar system's giant planets. Io, the innermost of the four big moons of Jupiter, for example, is the volcanically most active body in the solar system \citep{1979Sci...204..951S} and its primary source of internal energy is tidal heating. As the heat generated by internal bodily friction comes at the expense of a loss of rotational and/or orbital energy, strictly speaking the source of the internal heat is actually external to the object that is heated (here: Io). Orbital perturbations from the other major satellites around Jupiter force Io on an eccentric orbit \citep{1979Sci...203..892P}. Io's global mean tidal energy flux is about $2\,{\rm W\,m}^{-2}$ \citep{2000Sci...288.1208S}. Another prominent example is Saturn's moon Enceladus, which shows tidally driven cryovolcanism \citep{2006Sci...311.1393P}. Similar to the case of Io, the reason for the ongoing tidal heating in Enceladus is its forced orbital eccentricity. The ultimate circularization of its orbit is inhibited by its gravitational interaction with another neighboring moon, in this case of Saturn's moon Dione \citep{Meyer2007}.

Although detailed modeling of the tides in the Hadean Earth-Moon system cannot be found in the literature, the nearby Moon must have strongly deformed the Earth from its equilibrium shape. The fast rotation of the Earth compared to the orbital motion of the Moon led to a time lag \citep[or a phase lag;][]{2009ApJ...698L..42G,2013ApJ...764...26E} between the line connecting the two centers of mass and the instantaneous orientation of the tidal bulge on Earth. This offset then led to friction inside the Earth, which triggered the tidal heating. Previous studies estimated that the contribution of tidal heating on the Earth's surface was $\lesssim~0.1\,{\rm W\,m}^{-2}$ about 100\,Myr after the Moon-forming impact, and therefore irrelevant for the long term global energy budget \citep{2007SSRv..129...35Z}. These estimates were based on a specific tidal model (the constant phase lag tidal model) \citep{1879_Darwin,1880RSPT..171..713D,MacDonald1964,2008CeMDA.101..171F} that implies a homogeneous composition of the Earth and a specific parameterization of the Earth's tidal dissipation, technically speaking the second degree tidal Love number ($k_2$) \citep{1909RSPSA..82...73L,1911spge.book.....L} and the tidal dissipation constant ($Q$) \citep{1966Icar....5..375G}. Here we show that the tidal heating rates in the early Earth might have been higher than previously thought.

\subsection{Energy balance in a gray atmosphere model}
\label{sec:balance}

We start our calculations by addressing the fact that the Earth's global mean effective surface temperature ($T_{\rm s}$) is given by the thermal equilibrium between the absorbed incoming solar radiation and other internal energy sources on the one hand, and the emitted infrared radiation on the other hand. With $F_{\rm em}$ as the emitted energy flux from Earth per unit surface area, we have \citep{stefan1879,1884AnP...258..291B,1901AnP...309..553P}

\begin{equation}\label{eq:T_s}
T_{\rm s} = {\Bigg (} \frac{F_{\rm em}}{\sigma_{\rm SB}} {\Bigg )}^{1/4} \ ,
\end{equation}

\noindent
where $\sigma_{\rm SB}~=~5.670~\times~10^{-8}\,{\rm W\,m}^{-2}\,{\rm K}^{-4}$ is the Stefan-Boltzmann constant. In modern Earth, $F_{\rm em}$ is dominated by the re-emission of the absorbed insolation,

\begin{equation} \label{eq:flux}
F_{\rm em} = \frac{L_\odot(t)}{4 \pi (1\,{\rm AU})^2} \frac{(1-\alpha) \epsilon }{f} {\Big (} 1 + \frac{3}{4}\tau {\Big )} \ ,
\end{equation}

\noindent
where $L_\odot$ is the time-dependent solar luminosity ($t$ is time), $\epsilon$ is the Earth's emissivity\footnote{Emissivity describes the ratio of the thermal radiation from a surface of a given temperature compared to the radiation from a black body surface at the same temperature.}, $\alpha$ is the Earth's Bond albedo, $f$ is an energy redistribution factor that accounts for the Earth's rotation, AU~=~150\,million km is the Sun-Earth distance, and $\tau$ is the atmospheric infrared gray opacity \citep{2012RvGeo..50.2006F}. In this equation, the Earth's atmosphere is assumed to be gray \citep{Emden1913}, that is, its radiative properties are independent of wavelength ($\lambda$). This approximation neglects the strong $\lambda$ variability of the solar flux reflection, absorption, and re-emission by the various atmospheric gaseous components, dust, and clouds \citep{1969Icar...10..290S,Wei2019}. The term ``gray'' thus does not refer to the color of the atmosphere but to the non-dependence of the model on the wavelength.

For modern Earth, a typical parameterization of Eq.~\eqref{eq:flux} uses $\alpha~=~0.3$ and $f=4$ \citep{2007A&A...476.1373S}. Moreover, the gray opacity in the visible regime of the electromagnetic spectrum (called the optical depth) has been measured at the Earth's surface as $\tau~\sim~0.35$ \citep{Terez2003}. For now, we take this value as a proxy for $\tau$, which is assumed to be independent of wavelength, in the infrared but we rectify this assumption in Sect.~\ref{sec:combi} and calibrate $\tau$ with the observed greenhouse warming. The Earth's emissivity can be approximated as $\epsilon~=~0.95$, between the values for water (0.96) and limestone (0.92). For modern values of the solar luminosity Eq.~\eqref{eq:flux} yields $T=-6.3^\circ$C. The deviation of $20.3^\circ$C from the actual value of $+14^\circ$C is due to the well-known $\lambda$ dependence of the atmospheric greenhouse effect -- mostly due to H$_2$O, CO$_2$, and CH$_4$ -- that is encapsulated in $\tau$. Obviously, the optical depth is not a good proxy for the infrared gray opacity and we return to this aspect in Sect.~\ref{sec:solar} and Fig.~\ref{fig:fys}. Nevertheless, Eq.~\eqref{eq:flux} is a much better approximation than that of an airless Earth, which has a theoretical global mean equilibrium temperature of $-18^\circ$C.

Early in the Earth's history, during as much as the first 1\,Gyr after formation, the solar luminosity was only about 70\,\% of its current value, for which Eqs.~\eqref{eq:T_s} and \eqref{eq:flux} predict $T_{\rm s}=-29.1^\circ$C. Even additional heating from a greenhouse effect of about $+20.3^\circ$C, as observed today, would be insufficient by far to prevent global freezing of the early Earth. More complex models that include both a warming greenhouse effect and an ice-albedo effect, which dramatically cools the planet as soon as it starts to have ice sheets, suggests that the Earth should have been even much cooler, raising the question of the yet unsolved early-faint-sun or faint-young-sun paradox \citep{1997Sci...276.1217S}. That said, the composition of the early Earth atmosphere was entirely different from its modern counterpart and significant amounts of CO$_2$, CH$_4$, and NH$_3$ could have further raised surface temperatures by several degrees Celsius. Nonetheless, atmospheric effects alone are unable to solve the faint-young-sun paradox \citep{1988SciAm.258b..90K,1997Sci...276.1217S}.

In the following, we use theoretical models to investigate the possible role of tidal heating in the Hadean and Archean Earth due to the young, nearby Moon. We use sedimentological and geochemical evidence for the presence of liquid water as benchmarks (see Sect.~\ref{sec:introduction}).

\begin{figure}
\centering
\includegraphics[width=0.8\linewidth]{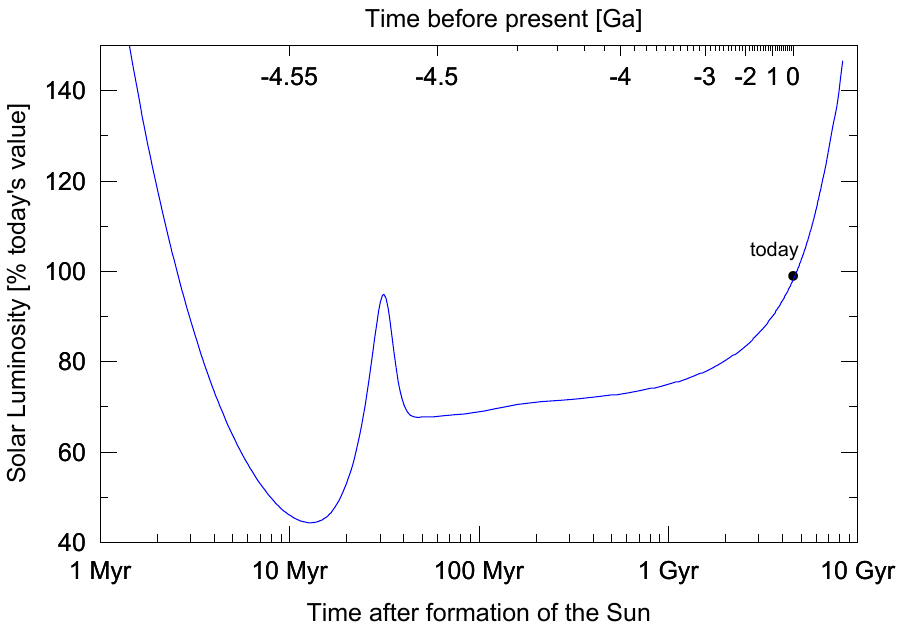}
\caption{Evolution of the solar luminosity as computed from a stellar evolution model \citep{2015A&A...577A..42B}. The age of the Sun of about 4.567\,Gyr is labeled with a black dot, at which time the model predicts a luminosity of 98\,\% of the actual observed value. The abscissa at the bottom shows the modeled time after formation of the Sun-like star. The abscissa at the top shows the time before present in units of gigayears ago (Ga).}
\label{fig:L_sun}
\end{figure}

\section{Evolution of the solar luminosity}
\label{sec:solar}

\begin{figure}
\centering
\includegraphics[width=0.8\linewidth]{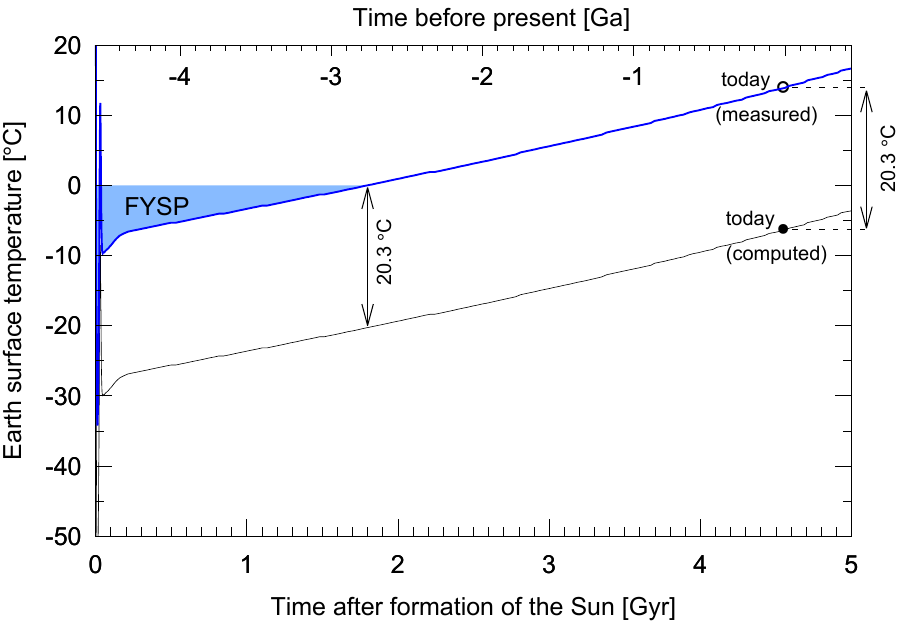}
\caption{Evolution of the early Earth's mean surface temperature under the assumption of a gray atmosphere, in which the wavelength dependence of the absorption of sunlight and re-emission in the infrared is neglected. The black thin line shows results as per Eqs.~\eqref{eq:T_s}-\eqref{eq:flux} with $\tau~=~0.35$ and using stellar evolution models \citep{2015A&A...577A..42B}. These models are multiplied by a factor of 1.02 to account for the offset between the predicted and the observed solar luminosity. The surface temperature computed for today is $-6.3^\circ$C. The thick blue line shows a model forced to fit the measured value of $+14^\circ$C, for which $\tau~=~0.925$ is required. The blue shaded area highlights the phase of sub-zero degree temperatures that cannot be explained by the additional heating of $+20.3^\circ$C from a greenhouse effect as on modern Earth. FYSP stands for faint-young-sun paradox.}
\label{fig:fys}
\end{figure}

Since the beginning, the energy budget of the Earth has been inherently liked with the luminosity of the Sun. The solar luminosity describes the electromagnetic energy output across all wavelengths and it currently amounts to $3.85~\times~10^{26}$\,W \citep{Chapman1997} with decadal fluctuations of the order of 0.1\,\% \citep{2013ARA&A..51..311S}. The solar luminosity cannot reliably be reconstructed even on a level of 1\,\% for more than the past few decades. Hence, we resort to stellar evolution models to reconstruct $L_\odot(t)$ throughout the Earth's history.

Figure~\ref{fig:L_sun} shows a pre-computed stellar evolution model of a Sun-like star \citep{2015A&A...577A..42B}. The track illustrates the initial settling of the Sun on the main sequence, a phase which is dominated by radial shrinking and conversion of potential energy into heat. After about 20\,Myr, the proton-proton nuclear reaction chain \citep{Bethe1939} kicks in, upon which the Sun almost doubles its energy output. After ongoing shrinking until about 40\,Myr into the life of the Sun, its luminosity settles at about 70\,\% of its modern value and then increases at a rate of a little less than 1\,\% per 100\,Myr for the next Gyr. At the current age of the Sun, marked by a black dot at 4.567\,Gyr along the bottom abscissa, the model predicts 98\,\% of the actual value of the solar power output. In the following, we compensate for this offset by multiplying the stellar evolution model by a factor of 1.02, which acts as a calibration to reproduce today's insolation.\footnote{Not shown in Fig.~\ref{fig:L_sun} is the variation of the solar spectral energy distribution, which might also have had a significant effect on the evolution of the Earth's atmospheric chemistry and therefore on the climate. To give just one example, the high-energy (short-wavelength, $\lambda<100$\,nm) radiation of the early Sun (${\sim}4$\,Ga and before) was about 100 to 1000 times stronger than it is today \citep{2005ApJ...622..680R,2020A&A...636A..83S}.}

In Fig.~\ref{fig:fys} we illustrate the resulting evolution of the Earth's global mean effective surface temperature as per Eqs.~\eqref{eq:T_s} and \eqref{eq:flux}. Different from Fig.~\ref{fig:L_sun}, which uses a logarithmic scaling along the abscissa to highlight the strong luminosity variations of the young Sun, Fig.~\ref{fig:fys} uses a linear scaling. The blue shaded area, present until 1.8\,Gyr after formation of the Sun ($\sim~2.8$\,Ga and before), marks the era of the early Earth, in which an additional heating of $+20.3^\circ$C by a modern Earth-like greenhouse effect cannot prevent the early Earth from becoming a global snowball. This time scale agrees with previous estimates \citep{1997Sci...276.1217S}. The minimum temperatures achieved in this model are as low as nearly $-50^\circ$C before the Sun reaches an age of 50\,Myr. In extreme scenarios, additional greenhouse warming from CO$_2$, CH$_4$, and NH$_3$ could have prevented the late phase of the faint-young-sun paradox near 2.2\,Gyr after formation of the Sun but not the first 700\,Myr (about 3.9\,Ga and before) with $T_{\rm s}$ as low as $-30^\circ$C as predicted by the gray atmosphere model. The black line in Fig.~\ref{fig:fys} is obtained by using $\tau=0.925$, which raises temperatures by $20.3^\circ$C and reproduces the global mean surface temperature on the modern Earth.

\section{Tidal evolution}

\subsection{Mechanical energy loss through tides}

The effect of tidal heating -- caused by the tidal forces from the young and nearby Moon, which probably formed as close as about $3.8\,R_\oplus$ \citep{Canup2004} following a giant impact  $69\,(\pm10)$\,Myr after the beginning of the solar system \citep{Maltese2019} -- on the climate of the early Earth has hitherto not been studied in detail (but see \citealt{2020GeoRL..4785746B} for tidal simulations of the Earth-Moon system under possible Archean-like continental distributions). It has been suggested that tidal heating in the Earth has dropped below $100\,{\rm W\,m}^{-2}$ within a few Myr after the Moon forming event and below $0.1\,{\rm W\,m}^{-2}$ within 100\,Myr \citep{2007SSRv..129...35Z}. These estimates were based on parameterized models of tidal equilibrium theory with a fixed second order tidal Love number ($k_2$) and constant tidal dissipation factor ($Q$) for the Earth. It is well-known, however, that the tidal dissipation of viscous objects, such as the partly molten early Earth, is strongly dependent on the frequency of the tide-raising potential \citep{2009ApJ...698L..42G}, i.e. the Keplerian orbital frequency of the Moon in the reference frame rotating with the Earth. The feedback mechanism between the warming effect of tidal heating on the melt fraction of the early Earth's mantle and the resulting change of the efficiency of tidal heating, which have been studied for extrasolar Earth-like planets \citep{2009ApJ...707.1000H,2014ApJ...789...30H}, have also not been studied in the young Earth-Moon system so far.

We do not consider these feedback mechanisms or the frequency dependence of the tidal $Q$ factor in this report either. Instead, our point is that previous calculations underestimated the effect of the enormous transfer of angular momentum from the fast-spinning early Earth to the orbit of the Moon (and the concomitant strong tidal heating) after the Moon-forming impact. In principle, this mechanism has been known for a long time, both from theory \citep{1980GeoJ...61..573W,1982GeoJ...70..261W} and from the interpretation of a banded iron formation in Australia \citep{1986Natur.320..600W}. Only recent advances in computer simulations of the actual impact scenario, however, suggested that the post-impact Earth had an extremely short rotation period, possibly near 2.2\,hr \citep{Canup2012}, which is several times faster than previously assumed. As we demonstrate below, the resulting total amount of the dissipated tidal energy could have had a significant effect on the early Earth's energy budget.

To set the stage for our tidal heating calculations, we compute the total rotational and orbital energy in the Earth-Moon system directly after formation and today. Since the Earth-Moon system does not lose significant portions of its mechanical energy through other mechanisms than tidal dissipation, we can safely assume that the difference between the modern and early state of the system has been dissipated through tides almost exclusively inside the Earth.\footnote{Tides raised on the Earth by the Sun also contribute to the energy dissipation in the Earth's rotation, but solar tides were only on a percent level since the formation of the Earth-Moon system \citep{2001Icar..150..288C}.} We assume that both the Earth and the Moon are solid, homogeneous spheres with masses of $M_\oplus~=~5.9736~{\times}~10^{24}$\,kg and $M_{\rm \leftmoon}~=~7.3477~{\times}~10^{22}$\,kg and with radii of $R_\oplus~=~6378$\,km and $R_{\rm \leftmoon}~=~1737$\,km, respectively. Generally, the rotational energy of a solid sphere is $E_{\rm rot}~=~1/2 \, I \omega^2$, where $I~=~2/5 \ M R^2$ is the moment of inertia of a sphere with mass $M$ and radius $R$, and $\omega~=~v_{\rm rot}/R$ is the angular velocity. The equatorial rotational speed $v_{\rm rot}~=~2\pi R/P_{\rm rot}$ can be calculated using the rotational period $P_{\rm rot}$. As a consequence,

\begin{equation} \label{eq:E_rot}
E_{\rm rot} = \frac{1}{5} M {\Big (} \frac{2 \pi R}{P_{\rm rot}} {\Big )}^2 \ .
\end{equation}

In addition to the rotational energy, there is (negative) orbital energy in the system, which is mostly stored in the orbital motion of the Moon and given as

\begin{equation} \label{eq:E_orb}
E_{\rm orb,\leftmoon} = - \frac{1}{2} M_{\rm \leftmoon} {\Big (} \frac{2 \pi a}{P_{\rm orb}} {\Big )}^2 \ ,
\end{equation}

\noindent
where $a$ is the Moon's orbital semimajor axis on its Keplerian orbit around the Earth and $P_{\rm orb}~=~2\pi \sqrt{a^3/(G(M_\oplus+M_{\rm \leftmoon}))}$ is its orbital period, as stated by Newton's derivation of Kepler's third law of motion \citep{1619hmlq.book.....K}. 
For the initial post-impact state of the system we use $a'~=~3.8\,R_\oplus$, which gives $P_{\rm orb}'~=~10.4$\,hr.\footnote{Here and in the following, quantities referring to the early Earth-Moon system are labeled with a prime $(')$.} Today, $a~=~60.3\,R_\oplus$ and $P_{\rm orb}~=~27.32$\,d. The total energy of the early Earth-Moon system and of the modern Earth-Moon system then is

\begin{align} \label{eq:dissipated_erl}  \nonumber 
E_{\rm tot}' = & \ E_{\rm rot,\oplus}' + E_{\rm rot,\leftmoon}' + E_{\rm orb,\leftmoon}'\\ \nonumber
                  = & \  3.06~{\times}~10^{31}\,{\rm J} + 1.27~{\times}~10^{27}\,{\rm J} - 6.17~{\times}~10^{29}\,{\rm J}  \\
                  = & \  3.00~{\times}~10^{31}\,{\rm J}  \\ \nonumber
E_{\rm tot} = & \ E_{\rm rot,\oplus} + E_{\rm rot,\leftmoon} + E_{\rm orb,\leftmoon} \\ \nonumber
                  = & \  2.57~{\times}~10^{29}\,{\rm J} + 3.14~{\times}~10^{23}\,{\rm J} - 3.85~{\times}~10^{28}\,{\rm J} \\
                  = & \  2.19~{\times}~10^{29}\,{\rm J}\label{eq:dissipated_mod} \ .
\end{align}

\noindent
As a consequence, the change of the total energy in the system is $3~{\times}~10^{31}\,{\rm J} - 2.19~{\times}~10^{29}\,{\rm J} = 2.98~{\times}~10^{31}\,{\rm J}$. In other words, the amount of mechanical energy stored in the modern Earth-Moon system is about 0.7\,\% of the initial amount. 99.3\,\% of the initial amount of mechanical energy has been dissipated as heat through the Earth's surface in its $4.5$\,Gyr history. This amount of energy is comparable to the modern solar energy output within one day.

\subsection{Tidal heating}
\label{sec:tidal}

\subsubsection{Ad hoc tidal model}

The tidal evolution of the Earth-Moon system has initially been described in a rigorous mathematical framework by G.~H.~Darwin \citep{1880RSPT..171..713D}, who developed an equilibrium tide model. This model assumes that the gravitational force of the tide raiser (here the Moon) elongates the perturbed body (here the Earth) and that this distortion is slightly misaligned to the line connecting the two centers of mass. This misalignment is due to dissipative processes within the deformed body, that is, currently mostly in the Earth's oceans, and it leads to a secular evolution of the orbit and spin angular momenta \citep{1978tfer.conf...62Z}.

The resulting transfer of angular momentum, mostly between the Earth's rotation and the orbit of the Moon, leads to an increase of $a$ and, hence, to a decrease of $\dot{E'}_{\rm tid}$ on a time scale of 10\,Myr to 100\,Myr, the details of which are in fact unknown.
Detailed modeling of the tidal heating evolution in the early Earth would require a better knowledge of the planetary structure, in particular of the thickness and temperature of a possible mantle in which most of the tidal heating is supposed to occur \citep{2009ApJ...707.1000H}. Our knowledge of the internal evolution of the early Earth after the Moon forming giant impact, however, is far from being complete. In fact, these uncertainties in the parameterization of the model would dominate the (unknown and so far untestable) errors in the results of a detailed modeling of the orbital evolution in the early Earth-Moon system.

Because of these uncertainties, we start our reflections on the tidal heating with a simple but not necessarily implausible approach and characterize the evolution of the tidal heating rate by a characteristic tidal heat flux ($F_{\rm t,0}$) and a characteristic time scale for the tidal evolution ($t_{\rm tid}$). Orbital evolution due to tides as well as the resulting tidal heating often follow an exponential or near-exponential decay law \citep{2007SSRv..129...35Z}. We therefore parameterize the evolution of the tidal heat flux as

\begin{equation} \label{eq:tidal_heat}
F_{\rm t}(t) = F_{\rm t,0} \, e^{-t/t_{\rm tid} } \ .
\end{equation}

\noindent
As a boundary condition for Eq.~\eqref{eq:tidal_heat} the tidal surface heat flux integrated over the age of the Earth ($T$) and multiplied by the surface of the Earth must be equal to the difference between the amounts of mechanical energy in the early and modern Earth-Moon systems (Eqs.~\ref{eq:dissipated_erl}-\ref{eq:dissipated_mod}),

\begin{equation} \label{eq:boundary}
4{\pi}R_\oplus^2 \int\displaylimits_{0}^{T} {\rm d}t \ F_{\rm t}(t) = E_{\rm tot}'-E_{\rm tot}~=~2.98~{\times}~10^{31}\,{\rm J} \ .
\end{equation}

\subsubsection{Constant-phase-lag model}

The constant-phase-lag (CPL) model assumes that the phase (or angle) lag between the Earth's equilibrium tidal bulge and the line connecting the Earth-Moon barycenters is constant during the orbit \citep{2008CeMDA.101..171F,2009ApJ...698L..42G,2011A&A...528A..27H,2013ApJ...764...26E}. 

In the absence of significant eccentricity or obliquity, \citet{2017CeMDA.129..509B} finds that $Q_\oplus~=34.5$ results in an orbital evolution that places the Moon at the Earth's surface 4.5\,Ga. A similar approach by \citet{2015E&PSL.427...74Z} predicts $Q_\oplus\,{\sim}\,30$. Although the Moon did not form at the Earth's surface but instead near the Earth's Roche radius, beyond which tidal forces were too weak to prevent accretion of the Moon, \citet{2015E&PSL.427...74Z} show that this proxy results in initial tidal heating rates near $100\,{\rm W\,m}^{-2}$, at least for the first few million years post impact. For comparison, three-dimensional simulations of tidal dissipation in the early Earth oceans suggest that $Q_\oplus~\sim~1$ might, under certain circumstances, be plausible \citep{2020GeoRL..4785746B}. For our nominal parameterization of the CPL model, we chose $Q_\oplus=12$ as in modern Earth.\footnote{We also investigated alternative choices of $Q_\oplus$. Interestingly, whatever reasonable value we chose (1.2, 12, 120) the resulting track of the tidal heating rates remained almost unchanged. The reason for this is in the concomitant change of the rate of the orbital evolution: if $Q_\oplus$ is increased (decreased), the rate of the orbital expansion decreases (increases). As a result, for any point in time, the tidal heating would be higher (lower) due to the change in $Q_\oplus$, but this change is almost precisely counterbalance by the variation of the orbital semi-major axis. In fact, tidal heating changes only by a factor of a few even if we change $Q_\oplus$ by an order of magnitude.}

The rate of the tidal energy dissipation is calculated as the sum of the loss of orbital energy and rotational energy. To calculate the tidal energy dissipation in the early Earth as a function of time, we use the formulation of the CPL model by \citet{2011A&A...528A..27H}. Their set of differential equations for ${\rm d}a/{\rm d}t$, ${\rm d}e/{\rm d}t$, ${\rm d}\omega/{\rm d}t$, and ${\rm d}\psi/{\rm d}t$ ($\psi$ being the obliquity or angle between the spin normal and the orbital plane) allows us to calculate the spin-orbit evolution. We also use their equations for the change of the orbital energy ($\dot{E}_{\mathrm{orb},i}$), where $i~\in~\{\oplus, {\rm \leftmoon}\}$, and for the change of the rotational energy ($\dot{E}_{\mathrm{rot},i}$). The total energy released inside the Earth and the Moon then is

\begin{equation}\label{equ:E_tid_FM08}
\dot{E}_{\mathrm{tid},i} = - \ (\dot{E}_{\mathrm{orb},i} + \dot{E}_{\mathrm{rot},i}) > 0 \ ,
\end{equation}

\noindent
respectively, and the globally averaged mean flux of the tidal heating is

\begin{equation}
h_{{\rm tid},i} = \frac{\dot{E}_{\mathrm{tid},i}}{4 \pi R_i^2} \ .
\end{equation}

\subsubsection{Comparison of tidal models}

\begin{figure}
\centering
\includegraphics[width=0.8\linewidth]{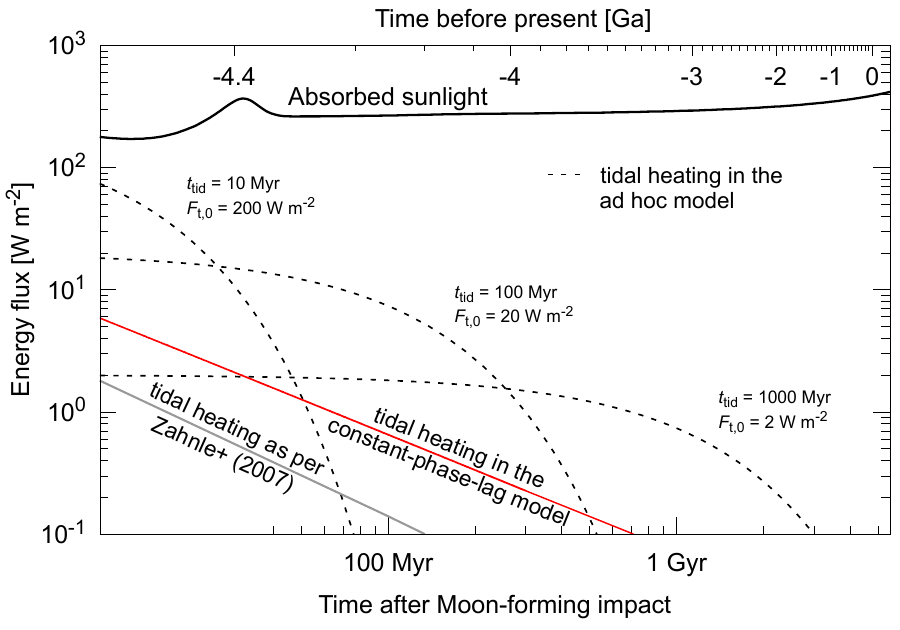}
\caption{Calculated tidal heat flux in the early Earth compared to the absorbed radiation from the young Sun (black solid line). Three different choices for the parameterization of the ad hoc model in Eq.~\eqref{eq:tidal_heat} result in the example tidal heat fluxes as labeled along the tracks (dashed lines). The evolution of the tidal heating rate as per the CPL model (red solid line) is based on numerical integrations of the orbital evolution of the early Earth-Moon system. Tidal heating rates from \citet{2007SSRv..129...35Z} are shown for comparison (gray solid line).}
\label{fig:tides}
\end{figure}

In Fig.~\ref{fig:tides} we show the tidal heat flux in the early Earth for three different choices of $t_{\rm tid}$ and $F_{\rm t,0}$ in Eq.~\eqref{eq:tidal_heat} that satisfy the energy loss criterium of Eq.~\eqref{eq:boundary} (dashed lines). All three choices are compared to the absorbed electromagnetic energy flux from the Sun as per Eqs.~\eqref{eq:T_s} and \eqref{eq:flux} and using pre-computed stellar evolution models \citep{2015A&A...577A..42B} (black solid line). We find that even the most extreme parameterization of the tidal heat flux with $t_{\rm tid}~=~10$\,Myr and $F_{\rm t,0}~=~200\,{\rm W\,m}^{-2}$ can barely compete with the solar absorbed flux on a time scale of $10$\,Myr, a time which would be measured after the Moon-forming impact. A more moderate and more reasonable choice of $t_{\rm tid}~=~100$\,Myr and $F_{\rm t,0}~=~20\,{\rm W\,m}^{-2}$ results in about $10\,{\rm W\,m}^{-2}$ of tidal surface heating for the first ${\sim}100$\,Myr after formation of the Earth and therefore could have been a significant source of energy during its early lifetime. This parameterization is used again as a reference tidal heat flux in Fig.~\ref{fig:fys_tides}. The resulting decay of tidal heating is in agreement with previous estimates \citep{2012RvGeo..50.2006F} of $\sim~0.02\,{\rm W\,m}^{-2}$ during the Archean (4\,Ga-2.5\,Ga). An even more moderate parameterization using $t_{\rm tid}~=~1000$\,Myr and $F_{\rm t,0}~=~2\,{\rm W\,m}^{-2}$ would keep tidal heating about two orders of magnitude smaller than the amount of absorbed sunlight.

We also compare the ad hoc tidal model to the tidal heating rates computed via numerical integration of the differential equations for the CPL model (red solid line). The shape of the curve of $h_{{\rm tid},i}(t)$ in the CPL model is remarkably different from the shape of the track in the ad hoc model. Any parameterization of the ad hoc model that we examine predicts stronger initial tidal heating than the CPL model. In turn, the CPL model provides more tidal heating in the long term, with over $1\,{\rm W\,m}^{-2}$ during the first 70\,Myr and up to $0.1\,{\rm W\,m}^{-2}$ during the first 700\,Myr. We have validated that the total energy dissipated through the Earth's surface is the same in all cases of the ad hoc model and the CPL model. Previously postulated tidal heating rates \citep{2007SSRv..129...35Z} are a factor of three to five smaller than the values that we compute using the CPL model. This is likely due to different initial conditions (in particular the rotation rate of the post-impact Earth) and different $Q_\oplus$ values.

Although all these tidal models produce tidal heat fluxes that are small compared to the solar energy input, even a conservative value of $F_{\rm t,0}~=~2\,{\rm W\,m}^{-2}$ is comparable to the global mean flux from tidal heating observed on Jupiter's volcanically active moon Io \citep{2000Sci...288.1208S}. It has been argued that this level of a surface heating through an internal heat source could trigger global volcanism on Earth-sized planets as well \citep{2009ApJ...700L..30B}. If this were the case, then even our most conservative choice of the parameterization of the evolution of tidal heating could have resulted in extreme geophysical processes on the early Earth that could possibly have produced strongly enhanced CO$_2$ outgassing and a substantial, long lasting greenhouse effect. In other words, beyond the direct heating effect of tidal friction, tides could have triggered further geophysical activities, which -- as a secondary effect -- could have heated the early Earth. Beyond that, the rapid resurfacing could have precluded the development of a biosphere \citep{2009ApJ...700L..30B}.

The comparison with the global mean internal heat emission of $2\,{\rm W\,m}^{-2}$ from the surface of the volcanically active Io introduces another interesting possibility. The widespread occurrence of ultramafic rocks such as komatiites before 3\,Ga has been interpreted as evidence for a higher geothermal gradient in the early Earth \citep{Nesbitt1982,Herzberg2010}. A detailed study of the convective heat and matter transport in the early Earth's mantle and crust are beyond the scope of this article. However, our results demonstrate the potential of tidal heating as a mechanism to explain, at least partly, the observations of two phenomena, that is, the faint-young-sun paradox and the enhanced internal heat flow in the Archean.

On the more speculative side, the tidal forces acting on the crust of the Archean Earth might have left scars in the geological record. The thousands of black chert veins of the Dresser Formation (North Pole region, Western Australia; see Fig~\ref{fig:chert_veins}a,b)  formed in a volcanic caldera system \citep{VanKranendonk2008} document intensive brittle deformation of the crust 3.49\,Ga ago. In the light of our findings, it is tempting to hypothesize that the obvious mechanical disruption of the crust was at least in parts caused by strong tidal forces imparted by the nearby early Moon. This would also be in good accordance with abundant carbonate-filled fractures as e.g. observed in the 3.47\,Gyr old Mt. Ada Basalt (see Fig~\ref{fig:chert_veins}c). Taken together, significantly higher tidal forces due to the much closer early Moon plausibly resulted in brittle deformation of the crust and enhanced geothermal activity in the Archean Earth.

\section{Combination of solar luminosity and tidal heating}
\label{sec:combi}

\begin{figure}
\centering
\includegraphics[width=0.8\linewidth]{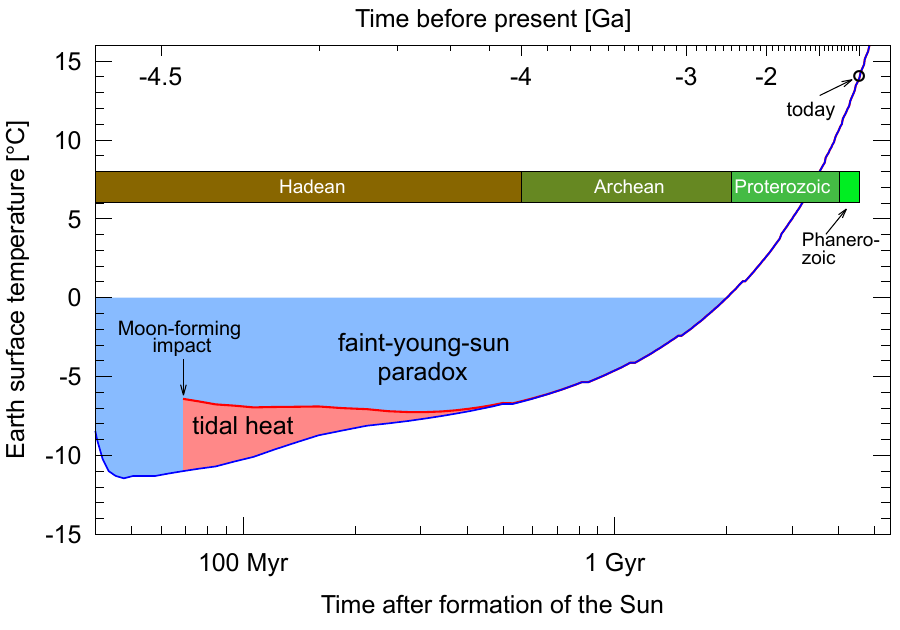}
\caption{Evolution of the early Earth's mean surface temperature using energy input from both insolation and tidal heat flux. The gray opacity of the Earth atmosphere is set to $\tau~=~0.925$ to force the atmosphere model to reproduce modern Earth temperatures. The blue line shows results as per Eqs.~\eqref{eq:T_s} and \eqref{eq:flux} and using stellar evolution models \citep{2015A&A...577A..42B} calibrated to reproduce the modern power output of the Sun. The blue shaded area highlights the phase of sub-zero temperatures. The red shaded area, the integral of which contains $2.98~{\times}~10^{31}\,{\rm J}$ as per Eq.~\eqref{eq:boundary}, illustrates an increase of the temperature due to tidal heating assuming $F_{\rm t,0}~=~20\,{\rm W\,m}^{-2}$ and $t_{\rm tid}~=~100$\,Myr in Eq.~\eqref{eq:tidal_heat}. Geological ages are highlighted with horizontal bars. In summary, tidal heating alone cannot possibly solve the faint-young-sun paradox. Nevertheless, it could have been an important ($>1^\circ$C) compensation of the low solar luminosity in the first ${\sim}~220$\,Myr (before about 4.35\,Ga) to prevent a global snowball Earth.}
\label{fig:fys_tides}
\end{figure}

Equipped with Eq.~\eqref{eq:flux} for the globally averaged insolation absorbed by the Earth, a pre-computed track of the solar luminosity to feed into Eq.~\eqref{eq:flux}, and an evolutionary model for the Earth's tidal heat flux as per Eq.~\eqref{eq:tidal_heat}, we can now set $F_{\rm em}=F_\odot+F_{\rm t}$ in Eq.~\eqref{eq:T_s} and re-evaluate the evolution of the Earth's global mean surface temperature.

We calibrate Eq.~\eqref{eq:flux} to reproduce the modern Earth value of $T_{\rm s}=+14^\circ$C by setting the infrared gray opacity to $\tau=0.925$. This approach can be seen as forcing the gray atmosphere model to encapsulate all the complex radiative properties of the Earth's atmospheric gases in the free parameter $\tau$. We also choose $F_{\rm t,0}~=~20\,{\rm W\,m}^{-2}$ and $t_{\rm tid}=100$\,Myr in Eq.~\eqref{eq:tidal_heat}, a parameterization that gives a total amount of dissipated tidal energy through the Earth's surface that is compatible with the observations from orbital mechanics (Eqs.~\ref{eq:dissipated_erl}-\ref{eq:dissipated_mod}).

In Fig.~\ref{fig:fys_tides} we plot $T_{\rm s}(t)$ for this parameterization of the model. The blue line at the bottom refers to the gray atmosphere model (forced to reproduce modern Earth surface temperatures) without the additional contribution of tidal heating. The minimum temperatures reached in this model are $-11.5^\circ$C at 50\,Myr after formation of the Sun.

The red line in Fig.~\ref{fig:fys_tides} refers to Earth's mean surface temperature based on the sum of the tidal plus solar absorbed flux on Earth. The contribution from tidal heating on the early Earth's surface temperature is highlighted by the red shaded area. In this model, the Moon-forming impact \citep[set to 69\,Myr after the formation of the Sun;][]{Maltese2019} results in an addition of tidal heat flux, which keeps temperatures near $-7^\circ$C instead of the $-11.5^\circ$C mentioned above. In the following, the tidal heat flux decreases while the solar luminosity increases by comparable fraction. As a consequence, the Earth's surface temperature remains almost constant near $-7^\circ$C. After this phase, tidal heating vanishes, whereas the solar luminosity continues to climb. The temperature offset triggered by tidal heating is up to about $+5^\circ$C within the first 10\,Myr after the Moon-forming impact and $>1^\circ$C for the first $\sim150$\,Myr of the newly formed Earth-Moon system. Assuming a Moon-forming impact 69\,Myr after formation of the Sun, our results imply that tidal heating could have been significant for the maintenance of liquid surface water up until ${\sim}220$\,Myr after the formation of the Sun, that is, until about 4.35\,Ga (see Fig.~\ref{fig:fys_tides}). The periods of the Hadean (4.56\,Ga--4\,Ga), Archean (4\,Ga--2.5\,Ga), Proterozoic (2.5\,Ga--541\,Ma), and Phanerozoic (541\,Ma--today) geological ages are also shown in Fig.~\ref{fig:fys_tides}. This comparison shows that tidal heating in the Earth due to the nearby Moon was only relevant in the Hadean, at least for this choice of a paramterization of our model. 

We have also explored other reasonable parameterizations of Eq.~\eqref{eq:tidal_heat} and found that, in no case, tidal heating alone could contribute the required amount of heat to bypass the faint-young-sun paradox and prevent a global snowball Earth. Either tidal heating is short ($t_{\rm tid}~{\lesssim}~10$\,Myr) and extreme ($F_{\rm t,0}\,{\sim}\,200\,{\rm W\,m}^{-2}$), in which case it is sufficiently strong to push the global mean temperature of the early Earth above $0^\circ$C -- but only for a few Myr and therefore much too short. Or tidal heating acts on a longer time scale ($t_{\rm tid}~{\gtrsim}~100$\,Myr) and with a more moderate magnitude ($F_{\rm t,0}~\sim~20\,{\rm W\,m}^{-2}$), but then it never lifts $T_{\rm s}$ above $0^\circ$C.

Irrespective of the details of the actual evolution of tidal heating, the overall dissipation of the mechanical energy budget (Sect.~\ref{sec:tidal}) suggests that this mechanism was not negligible in the early Earth. On the contrary, we demonstrate that tidal heating factored significantly into the heat budget of early Earth, likely contributing to the prevention of a global snowball state.

\section{Conclusions}

We have shown that the modern Earth-Moon system only contains about 1\,\% of its initial spin-orbit energy budget, when the Earth was a fast rotator and the Moon was in an extremely tight orbit. Most of the energy has been dissipated through tides, and we show that the resulting tidal heating may have had a much larger effect on the Earth's mean surface temperature than previously thought. Our combination of a gray atmosphere model with an \textit{ad hoc} model for the evolution of the Earth's tidal heating rate, which is compatible with the total energy released in the Earth-Moon system, predicts that tidal heating may have contributed a heating of several degrees Celsius within the first ${\sim}100$\,Myr of the life of the Earth.

Although tidal heating in the Earth from a closer Moon cannot possibly solve the faint-young-sun paradox alone, our results suggest that it could have played an important role in maintaining liquid water at the surface. As a bonus, tidal heating as a geothermal heat source might have helped to sustain enhanced mantle temperatures, for instance by driving hydrothermal fluid circulation in early Earth's crust. This interpretation would be compatible with the previously stated ``hydrothermal pump'' concept, which asks for an internal heat source and strong forces in the crust of the Archean Earth to explain the presence of 3.49\,Gyr old black chert veins of the Dresser Formation (Western Australia). A unified model that includes geophysical, atmospheric, tidal, and astrophysical effects could be able to resolve the longstanding faint-young-sun paradox.\\

\begin{acknowledgements}
The authors are thankful to Rory Barnes and an anonymous referee for their helpful comments on the manuscript.
\end{acknowledgements}

%

\noindent {\bf Funding.} RH and LG acknowledge support from the German Aerospace Center (Deutsches Zentrum f\"ur Luft- und Raumfahrt) under PLATO Data Center grant 50OO1501. JPD and JR acknowledge support from the DFG SPP 1833 ``Building a Habitable Earth'' (DU\,1450/3-1, DU\,1450/3-2, and RE\,665/42-2).\\

\noindent {\bf Conflict of interest.} The authors declare that they have no conflict of interest.\\

\noindent {\bf Availability of data and material.} Not applicable.\\

\noindent {\bf Code availability.} The {\tt gnuplot} scripts used to plots Figs.~\ref{fig:L_sun}-\ref{fig:fys_tides} can be requested via e-mail from author RH (\href{mailto:heller@mps.mpg.de}{heller@mps.mpg.de}).

\bibliographystyle{spbasic}      
\bibliography{references}   

\end{document}